\documentclass{article}
\usepackage{spconf,amsmath,graphicx}
\usepackage{adjustbox}
\usepackage{cite}
\usepackage{hyperref}

\usepackage{caption}
\usepackage{amsfonts}
\title{Ambisonics Networks - The Effect Of Radial Functions Regularization }
\name{Bar Shaybet\textsuperscript{1}, Anurag Kumar\textsuperscript{2}, Vladimir Tourbabin\textsuperscript{2} and Boaz Rafaely\textsuperscript{1}}
\address{\textsuperscript{1}School of Electrical and Computer Engineering, Ben-Gurion University of the Negev
\\
\textsuperscript{2} Reality Labs Research @ Meta
}

\begin{document}
%\ninept
%
\maketitle
\begin{abstract}
Ambisonics, a popular format of spatial audio, is the spherical harmonic (SH) representation of the plane wave density function of a sound field. 
Many algorithms operate in the SH domain and utilize the Ambisonics as their input signal. The process of encoding Ambisonics from a spherical microphone array involves dividing by the radial functions, which may amplify noise at low frequencies. This can be overcome by regularization, with the downside of introducing errors to the Ambisonics encoding. This paper aims to investigate the impact of different ways of regularization on Deep Neural Network (DNN) training and performance. Ideally, these networks should be robust to the way of regularization. Simulated data of a single speaker in a room and experimental data from the LOCATA challenge were used to evaluate this robustness on an example algorithm of speaker localization based on the direct-path dominance (DPD) test. Results show that performance may be sensitive to the way of regularization, and an informed approach is proposed and investigated, highlighting the importance of regularization information.
\end{abstract}
\begin{keywords}
Ambisonics, Spherical array, Radial function regularization, DPD
\end{keywords}
\section{Introduction}
\label{sec:intro}

% Spatial audio is a diverse field encompassing technologies and techniques that enable listeners to have immersive auditory experiences and perceive sound in three-dimensional space, akin to real-world environments. 
Spatial audio gained significant popularity in recent years due to advancements in headphone technology and the increasing prevalence of VR/AR headsets. In light of these new developments, Ambisonics, a popular spatial audio format, enables headphone listening with effective head tracking \cite{book_ambisonic_format}.
%as well as spatial audio recording devices, such as the Zoom \cite{zoom_recorder} and Eigenmike \cite{eigenmike}.
% Ambisonics, ideally, contains all the information necessary for regenerating a full sphere sound field, allowing the listener to hear and rotate a complete reconstruction of the acoustic scene.
One way to capture spatial sound is using spherical microphone array \cite{book_boaz}, which consists of microphones placed over the surface of a sphere, whether rigid or open. Due to the spherical shape, encoding the microphone signals into Ambisonics signals is readily supported \cite{PWD_boaz}. 
% By representing the sound field as a summation of plane waves, transformation of the sound field to the SH domain and the execution of Plane Wave Decomposition (PWD) \cite{PWD_boaz} extract the Ambisonics, amplitude representation of plane waves from various directions.

Spatial signal processing in the spherical harmonics (SH) domain using Ambisonics signals exists for wide variety of purposes such as direction of arrival estimation \cite{ambi_doa_rangeEstimation}, speech enhancement \cite{ambi_moti_speechEnhancment}, binaural reproduction \cite{ambi_binaural} and source separation \cite{ambi_algo_source_seperation}.
% Mathematically, Ambisonics consists of coefficients in the Spherical Harmonic (SH) domain. When the sound field is represented as a summation of plane waves, the transformation of the sound field to the SH domain and the execution of Plane Wave Decomposition (PWD) \cite{PWD_boaz} extract the Ambisonics, amplitude representation of plane waves from various directions.
% Spatial signal processing in the SH domain is convenient and exists for wide variety of purposes such as direction of arrival estimation, speech enhancement and binaural reproduction \cite{ambi_moti_speechEnhancment,ambi_doa_rangeEstimation,ambi_coding,ambi_algo_source_seperation,ambi_binaural}.
% One way to capture spatial sound is using Spherical Microphone Array (SMA) \cite{book_boaz}, microphones that lays on sphere surface, rigid or open. Due to the spherical array shape, it is comparatively convenient to perform SH transformation and PWD to calculate the Ambisonics.
However, the encoding of Ambisonics signals is not error-free due to the limited number of microphones \cite{book_boaz} and the effect of measurement noise \cite{book_ville_pullki}. Additionally, Ambisonics encoding using the Plane-Wave Decomposition (PWD) \cite{PWD_boaz}, involves a division by the array radial functions which may have a small magnitudes at low frequencies. The latter may lead to noise amplification at these frequencies. To overcome this limitation, several methods for regularization were proposed to offer a the trade-off between limiting the unwanted noise gain while avoiding loss of spatial information in the Ambisonics signal \cite{book_ville_pullki}.   

In the last decade, Deep Neural Networks (DNN) gained popularity across all fields, and in recent years algorithms based on Ambisonics signals and DNN were developed for a variety of tasks, such as speaker localization \cite{DPD_orel}, speech separation \cite{adrian_speechSeperation, ambi_algo_voice_seperation_DNN, ambi_algo_source_seperation_DNN} and enhancement \cite{ambi_algo_speach_enhancment}. The matter of regularization is often overlooked in these studies, while robustness of networks based on Ambisonics signals to different regularization techniques has yet to be explored to the best of our knowledge.   

In this paper, we study and demonstrate the potential sensitivity of Ambisonics neural networks to varying levels of regularization. Moreover, to overcome this sensitivity, we demonstrate that utilizing  the regularization information may be beneficial. 
This study presents two main contributions:
\begin{itemize}
    \item Demonstration using both simulated and measured data that the performance of networks based on Ambisonics signals may be compromised when employing Ambisonics encoding with different regularizations.
    \item Evidence that incorporating regularization information can enhance the performance of Ambisonics networks by increasing their robustness to varying regularization levels.
    % \item Demonstration that stronger regularization yields diminished performance of the DNN-DPD test, when the model was initially trained with minimal regularization strength.  
\end{itemize}

\section{Ambisonics Encoding}
\label{sec:Spherical_Harmonics_Encoding}
This section presents an overview of the methods used to encode Ambisonics from spherical array data using regularized PWD.
\subsection{Regularized PWD}
\label{sec:Regularized_PWD_Model }
The signal measured by a spherical microphone array in a sound field represented by Ambisonics signals and additive noise can be formulated in the SH domain in matrix form as follows \cite{book_boaz}:

\begin{equation}
\label{eq:pyba}
% p = YBa
\mathbf{p}(k, r, \Omega) = \mathbf{Y}(\Omega)\mathbf{B}(kr)\mathbf{a}(k) + \mathbf{n}(k)
\end{equation} %%% is it correct? 
\\
where \( \mathbf{p}(k, r, \Omega) = [p(k, r, \Omega_1), \ldots, p(k, r, \Omega_Q)]^T \) is a vector holding the $Q$ microphone signals, \(k\) is the wave number, \(r\) is the distance of the microphones from the center of the array and \( \Omega = [\Omega_1, \dots,\Omega_Q] = [(\theta_1, \phi_1), \dots, (\theta_q, \phi_q)]\) is the set of angles representing microphone positions, where \(\theta_q, \phi_q\) are elevation and azimuth respectively. 
\( \mathbf{Y}(\Omega) = [\mathbf{y}(\Omega_1),\mathbf{y}(\Omega_2), \ldots, \mathbf{y}(\Omega_Q)]^T \) is a \(Q \times (N+1)^2\) matrix where 
\( \mathbf{y}(\Omega_q) = [Y_0^0(\theta_q, \phi_q), Y_{1}^{-1}(\theta_q, \phi_q), \ldots, Y_N(\theta_q, \phi_q)] \).
\(Y^m_n(\theta, \phi)\)  is the SH function of order $n$ and degree $m$ where $-n \leq m \leq n$. 
$\mathbf{a}(k) = [a_{00}(k),\ldots,a_{NN}(k)]^T$ is an \((N+1)^2 \times 1 \) vector, holding the Ambisonics signals of order $N$.
$\mathbf{B}(kr)$ is a diagonal matrix, holding the radial functions related to the array, with \( \mathbf{B}(kr) = Diag([b_0(kr), \ldots, b_N(kr)]) \). The radial functions depend on the array configuration \cite{book_boaz} and for a rigid sphere are given by

\begin{equation}
\label{eq:spheres}
b_n(k r)=4 \pi i^n\left(j_n(k r)-\frac{j_n^{\prime}\left(k r_{\mathrm{s}}\right)}{h_n^{(2)^{\prime}}\left(k r_{\mathrm{s}}\right)} h_n^{(2)}(k r)\right)
\end{equation} 
where \( j(kr)\) and \(h^{(2)}(kr) \) are the spherical Bessel functions of the first kind and spherical Hankel functions of the second kind respectively. The operator \({}^\prime\) denotes derivative.

\(\mathbf{n}(k)\) is a \(Q \times 1\) vector representing additive microphone noise assumed white and i.i.d. with zero mean satisfying: $E\left[\mathbf{n}(k) \mathbf{n}^H(k)\right]=\sigma_n^2 \mathbf{I}$ where $\sigma_n^2$ is the variance of the noise.

Solving Eq. \eqref{eq:pyba} for $\mathbf{a}(k)$ by performing PWD, and omitting the matrix and vector arguments for simplicity,  we obtain \cite{book_ville_pullki}: 
\begin{equation}
\label{eq:solution}
\mathbf{\hat{a}} =  \mathbf{(YB)}^{\dagger}\mathbf{p} = \mathbf{a} + \mathbf{(YB)}^{\dagger}\mathbf{n}
\end{equation}
\\
% where $\mathbf{C}$ is the identity matrix when no regularization method is used.

 In this paper we assume matrix $\mathbf{B}$ for rigid sphere which is a common configuration. The radial functions may have low magnitudes or zero at the origin, leading to significant noise amplification as suggested by Eq. \eqref{eq:solution}. One approach to mitigate this noise gain is by employing regularized PWD \cite{radial_filters_design,radial_filters_design2} by incorporating regularization matrix $\mathbf{C}$ in the following manner:
 \begin{equation}
\label{eq:solutionC}
\hat{\mathbf{a}}_R =  \mathbf{C(YB)}^{\dagger}\mathbf{p} = \mathbf{Ca} + \mathbf{C(YB)}^{\dagger}\mathbf{n}
\end{equation}
\\
While noise can be suppressed by an appropriate choice of $\mathbf{C}$, the accuracy in estimating $\mathbf{a}$ may be degraded as $\mathbf{C(YB)}^{\dagger}$ may not equal the unit matrix. Various regularization matrices $\mathbf{C}$ have been proposed to control the noise gain at low frequencies while minimizing distortion.
 Examples for such regularization methods are the Delay and Sum PWD \cite{delay_and_sum} and the R-PWD \cite{book_ville_pullki}. Another regularization method is based on the Tikhonov function \cite{book_tikhonov}, a diagonal matrix \(\mathbf{C}\) having the following elements on the diagonal: 
\begin{equation}
\label{eq:Tikhonov}
c_n(k)=\frac{\left|b_n(k r)\right|^2}{\left|b_n(k r)\right|^2+\lambda^2}
\end{equation}
This particular function will be used in this work due to the effective parameterization by $\lambda$.
Note that \( \lambda = 0 \) leads to \(\mathbf{C}\) becoming the identity matrix, signifying the absence of regularization. On the other hand, when \( \lambda^2 \gg |b_n(k r)|^2 \) noise is significantly suppressed, yet the Ambisonics signal may suffer from substantial distortion.

\subsection{The effect of regularization}
Regularization controls the trade-off between noise amplification and distortion in the Ambisonics signal.
These two measures can be formulated as follow:

\begin{equation}
\label{eq:distortion}
DIST = \frac{||\mathbf{(C-I)a}||^2}{||\mathbf{a}||^2}  
\end{equation}
\begin{equation}
\label{eq:noise}
G_{noise} = \frac{\mathbb{E}
[||\mathbf{C(YB)}^{\dagger}\mathbf{n}||^2]}{\mathbb{E}[||\mathbf{n}||^2]}  
\end{equation}
where $\mathbf{I}$ is the identity matrix.
% Equations \eqref{eq:distortion}, \eqref{eq:noise} represent the introduced distortion and noise amplification, respectively. Solving optimization problem to minimize both criteria was suggested \cite{book_ville_pullki} under an assumption of  a diffuse sound field. In the general case, the choice of regularization method depends on the task's prioritization, such as reducing noise or preserving spatial information, along with the specific acoustic SNR within the relevant temporal frequencies.
\begin{figure}[htbp]
    % \centering
    \includegraphics[width=0.5\textwidth]{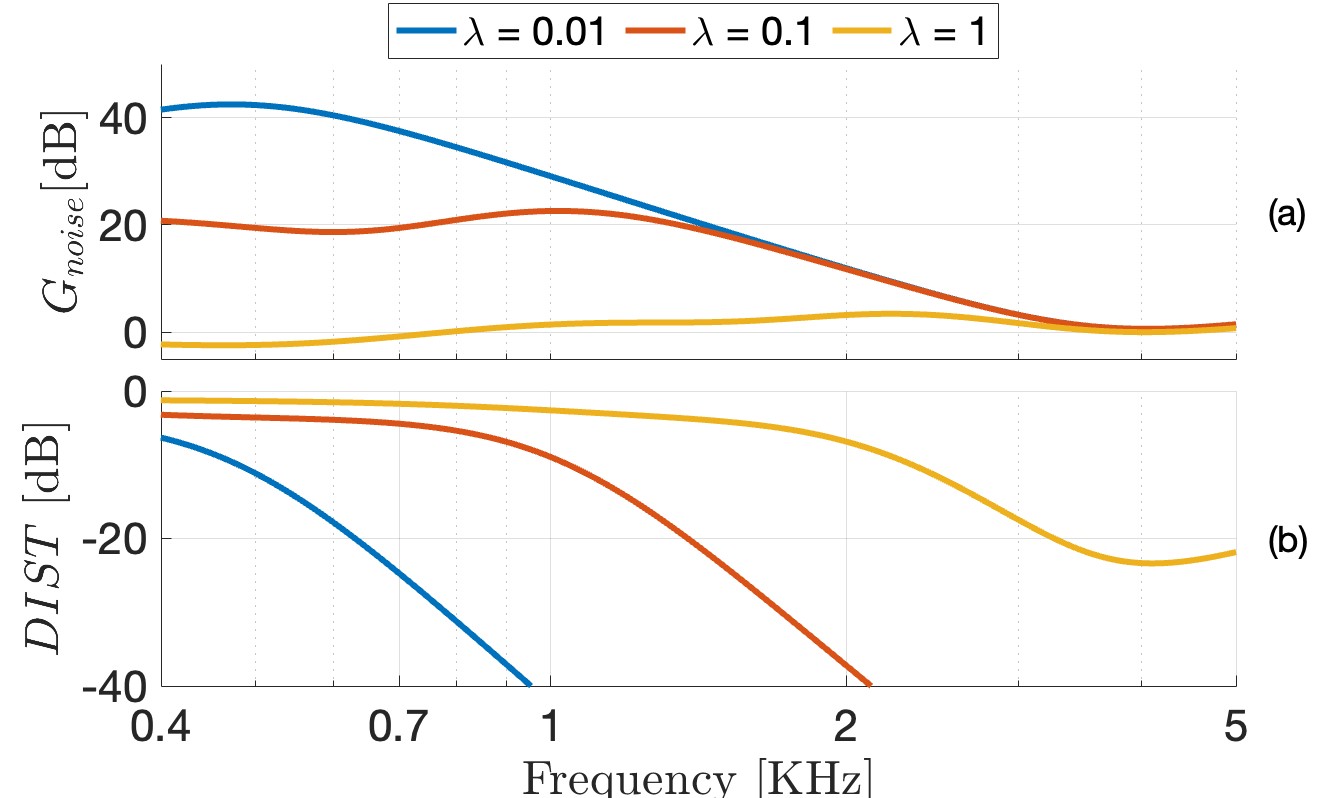}
    \caption{$(a)$ noise gain Gnoise, and $(b)$ distortion DIST as a function of frequency. In this example $\mathbf{a}$ represents a unit amplitude plane wave and $r=4.2$ cm.}
    \label{fig:Bs}
\end{figure}

As an illustrative example for the effect of regularization on these two measures, Fig \ref{fig:Bs} presents $DIST$ and $G_{noise}$ as defined in Eqs. \eqref{eq:distortion}, and \eqref{eq:noise}. In this example, $\mathbf{a}$ represents the SH coefficients of a single plane wave with an arrival direction of  $45^\circ$ in elevation and $20^\circ$ in azimuth. Also, array of radius $r=4.2$ cm was assumed. The figure clearly presents the trade-off between distortion and noise amplification through the choice of $\lambda$.
As the value of $\lambda$ increases, noise gain decreases while distortion increases, and vice versa.
Some methods employed an optimization to solve for the best trade-off \cite{book_ville_pullki}, but in the general case, the choice of regularization may depend on the specific task and the relative importance of noise gain and distortion. 
% Furthermore, It should be noted that Ambisonics of higher order, are impacted at higher frequencies.

\section{Speaker localization using the DNN-DPD algorithm}
\label{sec:DNN_DPD_algorithm}
To illustrate the sensitivity of an Ambisonics algorithm to different regularizations, this study assesses a speaker localization algorithm based on DNN with different levels of regularization. This section describes the details of the algorithm to provide context for the evaluation.

Direction Of Arrival (DOA) estimation algorithms such as MUSIC \cite{music} work very well in free field but tend to struggle in reverberant environments due to the correlated reflections. To address this challenge, the Direct Path Dominance (DPD) test was introduced \cite{DPD_Nadiri}. The objective of this algorithm is to select time-frequency bins associated with the direct sound, ensuring that DOA estimation is based on signal components that carry source direction information, i.e. the direct sound.

\begin{figure}[htbp]
    \vspace{-9pt}
    \centering
    \includegraphics[width=0.5\textwidth]{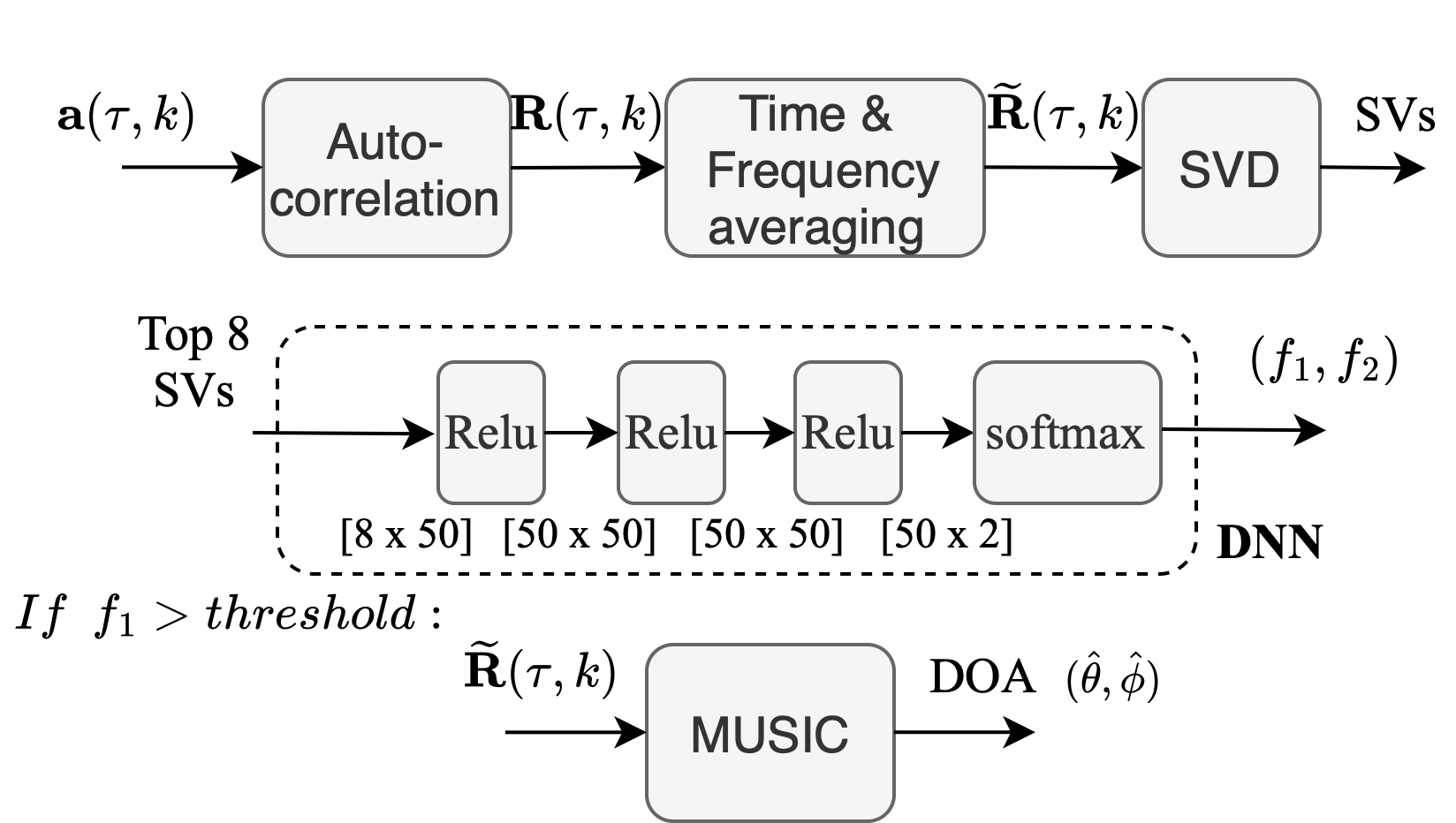}
    \caption{DNN-DPD based DOA estimation algorithm, showing the computation of the features, the details of the DNN-DPD block, followed by MUSIC. Network output, $(f_1, f_2),$ indicates the probability of the input to contain direct signal. $\hat{\theta}, \hat{\phi}$ are the estimated DOA of the speaker.} 
    \label{fig:net}
\end{figure}

Recently, the DPD test method has been extended to incorporate a neural network, called the DNN-DPD \cite{DPD_orel}, with input features based on  Ambisonics signals, and a direct sound classification output. First, STFT is applied to the Ambisonics signals with $\tau, k$ denoting time and frequency indices, respectively. Then, the Singular Values (SV) of the autocorrelation matrix, $\mathbf{R}(\tau,k)$, are computed from the Ambisonics signals, $\mathbf{a}(\tau,k)$. The DNN-DPD stage is followed by a DOA estimation stage based on MUSIC, as illustrated in Fig. \ref{fig:net}.

\section{Experimental Investigation}
\label{sec:Hypothesis}
This section presents an experimental investigation aiming to study the effect of Ambisonics regularization on the DNN-DPD algorithm. Furthermore, the importance of a priori regularization information is investigated by comparing informed and uninformed algorithms. 

\subsection{Setup}
The localization algorithm was trained using simulated data and tested using simulated and real data. For both cases, data of a single speaker in a room was used, captured with a spherical array equipped with 32 microphones, nearly uniformly distributed with a radius of 4.2 cm. Ambisonics was encoded to a 3rd order, while an STFT was applied to the time-domain signals with a window size of 512 and a 16kHz sampling frequency. Data in the frequency range [400Hz, 5000Hz] was used for processing.

\subsubsection{Simulated data}
\label{sec:Simulated_data}
Simulation data was generated using the Image Method \cite{image_method}, simulating the presence of a single speaker in a room. 
Speech signals were selected from TIMIT dataset \cite{timit}, 10 for training, and another 7 for testing, each signal 1 second long. Data was generated using various source and array locations as well as diverse room sizes (volumes of 24 to 480 $m^3$) and T60 (0.25 to 1.5 sec), similar to  \cite{DPD_orel}. White noise was added to the microphone signals, resulting in three levels of  Signal-to-Noise Ratios (SNR): [20dB, 40dB, 60dB] for the training data and 10dB for the test data. In total, 1008 seconds of audio were generated for training, which were divided to training set (80\%) and validation set (20\%). A binary label was computed for each time-frequency bin, assigning a positive value exclusively to bins with a Direct-to-Reverberant Ratio (DRR) greater than 0.7.
During training stage, the Ambisonics signals were calculated using the rigid body radial functions as in \eqref{eq:spheres} with $\mathbf{C}$ based on Tikhonov function as in \eqref{eq:Tikhonov}, with $\lambda = 0.001$. This value was selected to introduce
minimal regularization, following a similar approach as seen in other related papers \cite{adrian_speechSeperation}. 
An informed model, is also examined to show the importance of regularization information. This model utilizes the regularization information and its training and test data were generated using different $\lambda$ values [0.05, 0.5, 1.5] corresponding to SNR values [20dB, 10dB, 5dB] as suggested in \cite{book_ville_pullki}.  
\subsubsection{Measured data}
In addition to simulated data, measured data from the LOCATA challenge \cite{LOCATA} was also used for testing. We utilized 13 speech signals from task 1 of the challenge, each lasting 4 seconds, composed of a stationary speaker in the room which had a 0.5-second reverberation time. A regularization level of $\lambda=0.25$ was selected, according to the data SNR of 13 dB.

\subsection{Evaluation Methodology}
\label{sec:typestyle}

% Evaluation stage was done using different 7 speech signals from the same dataset of training but with different scenarios. Data permutation parameters for evaluation were chosen as it was suggested \cite{DPD_orel} where SNR chosen to be 10dB in our case. 
While training data was generated using one level of regularization, with the intention of observing the impact of different regularization levels, Ambisonics signals of the test data were calculated with varying \(\lambda\) values in the range of [0.01,1]. 

In addition, in order to show the importance of the regularization information, an informed algorithm was developed and evaluated, based on the algorithm described in Fig. \ref{fig:net} but differs in several aspects: first, diverse data was used to train the model as it described in \ref{sec:Simulated_data}. Then, the informed algorithm avoided the selection of highly distorted Ambisonics signals by weighting the DNN-DPD network output with a distortion-related weight:

\begin{equation}
W(k,\lambda) = 
\begin{cases}
1 - DIST & \text{if } DIST < 1, \\
0 & \text{otherwise}.
\end{cases}
\end{equation}
Finally, DOA estimation computed using the MUSIC algorithm. The DOA estimation error was finally calculated as following:
\\
\begin{equation}
\Theta_{E R R}=\frac{1}{L} \sum_{j=1}^{L} \operatorname{\sqrt{\frac{1}{2}\left((\hat{\theta}_j-\theta_j)^2+(\hat{\phi}_j-\phi_j)^2\right)}}
\end{equation}
\\
where \(L\) is the number of bins that passed the DNN-DPD test, \((\theta,\phi) \) are their true elevation and azimuth angles and \((\hat\theta,\hat\phi)\) are their estimated values.

%%%
\subsection{Results}
\label{sec:Results}
\begin{figure}[htbp]
    % \vspace{-8pt}
    \centering
    \includegraphics[width=0.5\textwidth]{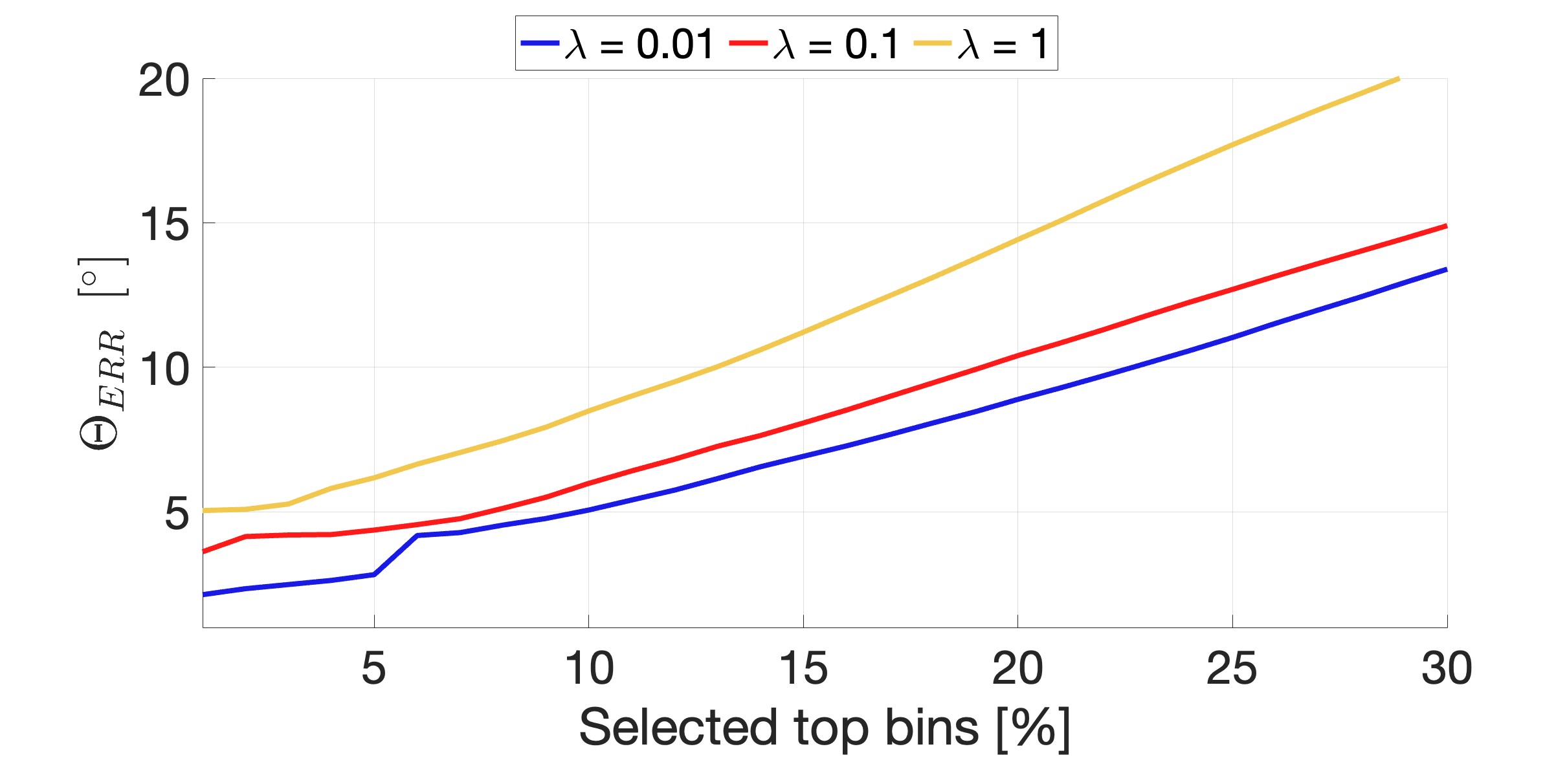}
    \caption{DOA error as function of the top-selected bins, simulated test data. Bins were selected according to the DNN-DPD network output $f_1$ in Fig. \ref{fig:net}}
    % The different Tikhonov \(\lambda\) values were tested separately.}
    \label{fig:TFerr}
\end{figure}

Fig. \ref{fig:TFerr} shows the DOA estimation error as a function of the top selected bins, for the uninformed algorithm for various regularized input.
Evidently, when stronger regularization is applied, the results deteriorate. The most substantial regularization exhibits the highest error, regardless of the number of bins considered.

\begin{figure}[htbp]
    \captionsetup{belowskip=-10pt}
    \centering
    \includegraphics[width=0.5\textwidth]{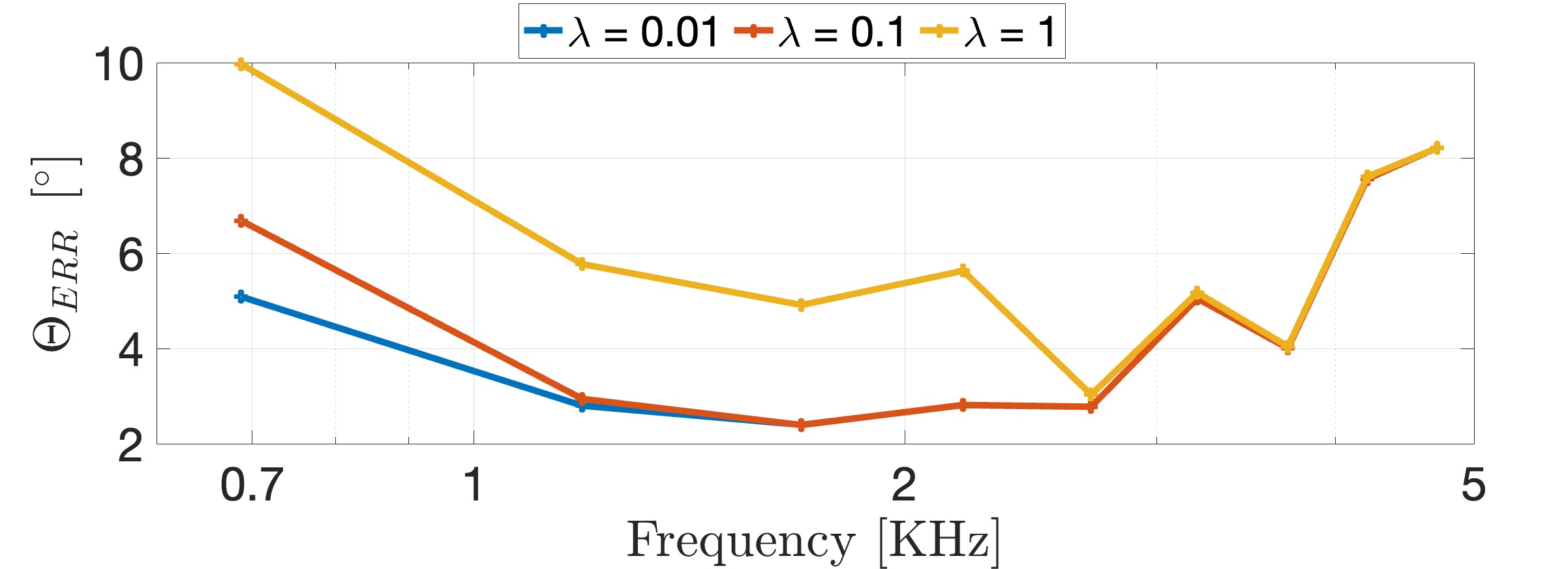}
    \caption{DOA error as function of frequency band for the test data. 5\% of top-scoring bins were selected from each frequency band. The markers represent the main frequency of each frequency band. }
    \label{fig:frequencyBand}
\end{figure}

Fig. \ref{fig:frequencyBand} presents the level of error per frequency band across various $\lambda$ values. For low frequencies the error varies significantly for different $\lambda$ values while the difference is less noticeable for higher frequencies.
This correspond with the behavior in Fig. \ref{fig:Bs}, indicating that frequencies with high distortion tend to introduce greater errors.

Fig. 5 compares the informed and uninformed algorithms, for both simulated data and the LOCATA data. The figure shows that the informed algorithm outperforms the
uninformed algorithm for the shown selected bins percentage, suggesting that regularization information can be useful and may improve algorithm performance when varying regularization levels are applied during Ambisonics encoding. 
% \begin{figure}[htbp]
%     \centering
%     \includegraphics[width=0.35\textwidth]{bands_snr10_top1.png}
%     \caption{Error per frequency band where SNR = 10 dB and 1\% top bins were selected.} %%% graph is relevant
%     \label{fig:TFerrBands}
% \end{figure}

\begin{figure}[htbp]
    \vspace{-6pt}
    \captionsetup{belowskip=-13pt,skip=6pt}
    \centering
    \includegraphics[width=0.50\textwidth]{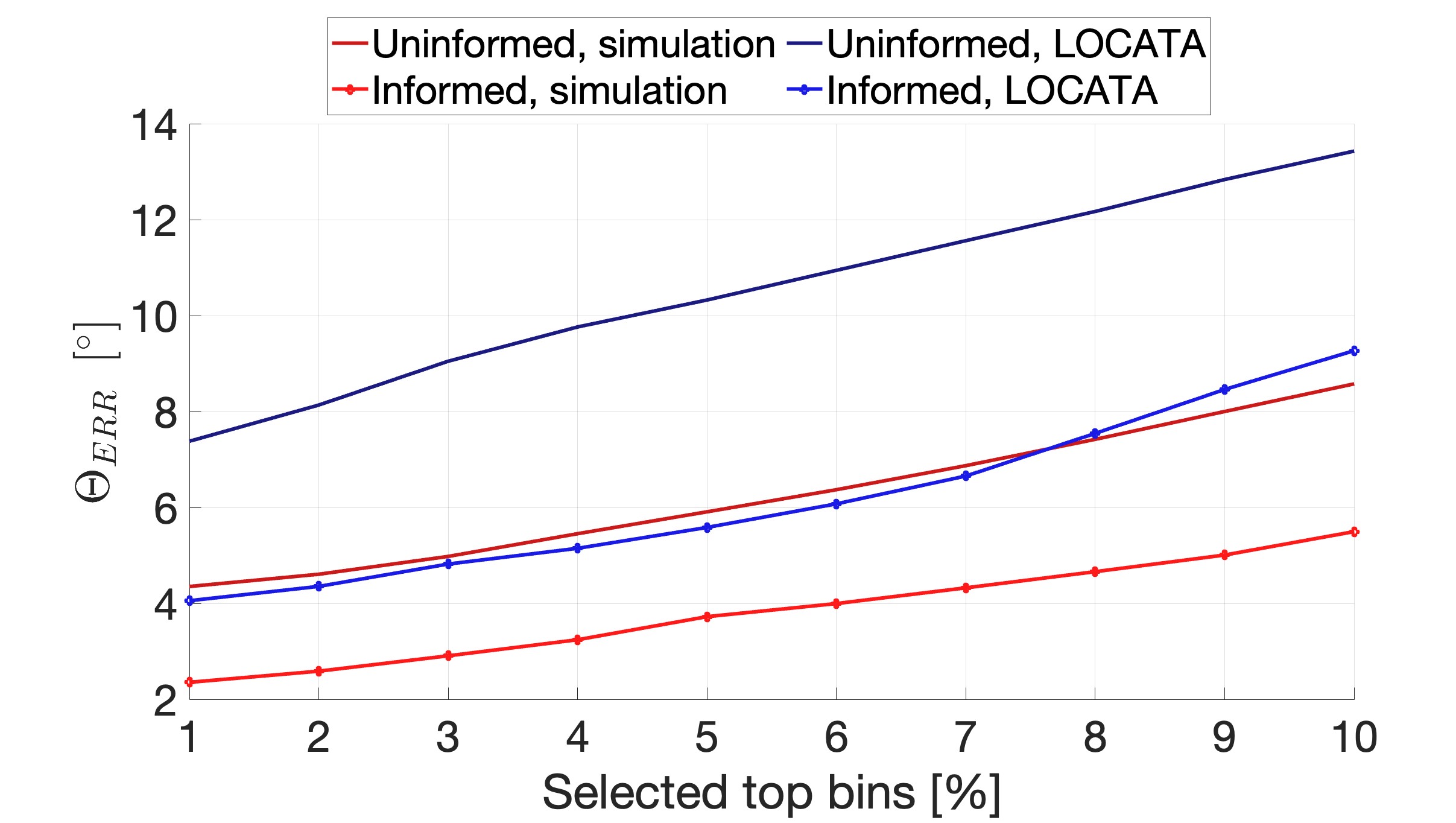}
    \caption{DOA estimation error as a function of selected top bins, for simulated and LOCATA data, with the informed and uninformed algorithms. }
    \label{fig:informed_tot}
\end{figure}

\section{Conclusions}
\label{sec:conclusion}

This study evaluates a localization Ambisonics algorithm performance across various levels of regularization. The uninformed algorithm which employed one regularization level during training, exhibits degraded performance when exposed to data with different regularization levels.
%; notably, stronger regularization yields worse results.
This highlights sensitivity of DNN algorithms to varying regularizations. An informed algorithm, with regularization information available, lead to improved performance, demonstrating the importance of informed regularization. Future work can study the generalization of the regularization-informed approach to other Ambisonics networks.

\bibliographystyle{IEEEbib}
% \bibliography{strings,refs}
\bibliography{sample}

\begin{thebibliography}{10}

\bibitem{book_ambisonic_format}
F.~Zotter and M.~Frank,
\newblock {\em Ambisonics: A Practical 3D Audio Theory for Recording, Studio
  Production, Sound Reinforcement, and Virtual Reality},
\newblock 01 2019.

\bibitem{book_boaz}
B.~Rafaely,
\newblock {\em Fundamentals of Spherical Array Processing},
\newblock Springer Topics in Signal Processing. Springer, Germany, second
  edition, 2019,
\newblock notValidatingIssn:1866-2609 ;.

\bibitem{PWD_boaz}
B.~Rafaely,
\newblock ``{Plane-wave decomposition of the sound field on a sphere by
  spherical convolution},''
\newblock {\em The Journal of the Acoustical Society of America}, vol. 116, no.
  4, pp. 2149--2157, 10 2004.

\bibitem{ambi_doa_rangeEstimation}
J.~Daniel and S.~Kitic,
\newblock ``Echo-enabled direction-of-arrival and range estimation of a mobile
  source in ambisonic domain,''
\newblock in {\em 2022 30th European Signal Processing Conference (EUSIPCO)},
  2022, pp. 852--856.

\bibitem{ambi_moti_speechEnhancment}
M.~Lugasi and B.~Rafaely,
\newblock ``Speech enhancement using masking for binaural reproduction of
  ambisonics signals,''
\newblock {\em IEEE/ACM Transactions on Audio, Speech, and Language
  Processing}, vol. 28, pp. 1767--1777, 2020.

\bibitem{ambi_binaural}
N.~R. Shabtai and B.~Rafaely,
\newblock ``Binaural sound reproduction beamforming using spherical microphone
  arrays,''
\newblock in {\em 2013 IEEE International Conference on Acoustics, Speech and
  Signal Processing}, 2013, pp. 101--105.

\bibitem{ambi_algo_source_seperation}
M.~Hafsati, N.~Epain, R.~Gribonvala, and N.~Bertin,
\newblock ``Sound source separation in the higher order ambisonics domain,''
\newblock 07 2019.

\bibitem{book_ville_pullki}
V.~Pulkki, S.~Delikaris-Manias, and A.~Politis, Eds.,
\newblock {\em Parametric Time-Frequency Domain Spatial Audio}, chapter~2,
\newblock Wiley-Blackwell, United States, Dec. 2017.

\bibitem{DPD_orel}
O.~Ben Zaken, B.~Rafaely, A.~Kumar, and V.~Tourbabin,
\newblock ``Direction of arrival estimation for reverberant speech based on
  neural networks and the direct-path dominance test,''
\newblock in {\em 2022 International Workshop on Acoustic Signal Enhancement
  (IWAENC)}, 2022, pp. 1--5.

\bibitem{adrian_speechSeperation}
A.~Herzog, S.~R. Chetupalli, and E.~A.~P. Habets,
\newblock ``{AmbiSep}: {A}mbisonic-to-{A}mbisonic reverberant speech separation
  using transformer networks,''
\newblock Bamberg, Germany, Sept. 2022.

\bibitem{ambi_algo_voice_seperation_DNN}
A.~Muñoz-Montoro, J.~Carabias-Orti, and P.~Vera-Candeas,
\newblock ``Ambisonics domain singing voice separation combining deep neural
  network and direction aware multichannel nmf,''
\newblock in {\em 2021 IEEE 23rd International Workshop on Multimedia Signal
  Processing (MMSP)}, 2021, pp. 1--6.

\bibitem{ambi_algo_source_seperation_DNN}
F.~Lluís, N.~Meyer-Kahlen, V.~Chatziioannou, and A.~Hofmann,
\newblock ``Direction specific ambisonics source separation with end-to-end
  deep learning,''
\newblock {\em Acta Acustica}, vol. 7, 06 2023.

\bibitem{ambi_algo_speach_enhancment}
A.~Bosca, A.~Guérin, L.~Perotin, and S.~Kitić,
\newblock ``Dilated u-net based approach for multichannel speech enhancement
  from first-order ambisonics recordings,''
\newblock in {\em 2020 28th European Signal Processing Conference (EUSIPCO)},
  2021, pp. 216--220.

\bibitem{radial_filters_design}
N.~Hahn and S.~Spors,
\newblock ``Further investigations on the design of radial filters for the
  driving functions of near-field compensated higher-order ambisonics,''
\newblock in {\em Audio Engineering Society Convention 142}, May 2017.

\bibitem{radial_filters_design2}
S.~L{\"o}sler and F.~Zotter,
\newblock ``Comprehensive radial filter design for practical higher-order
  ambisonic recording,''
\newblock {\em Fortschritte der Akustik, DAGA}, , no. 1, pp. 452--455, 2015.

\bibitem{delay_and_sum}
S.~Spors, H.~Wierstorf, and M.~Geier,
\newblock ``Comparison of modal versus delay-and-sum beamforming in the context
  of data-based binaural synthesis,''
\newblock 04 2012.

\bibitem{book_tikhonov}
A.~Tikhonov and V.~IA. Arsenin,
\newblock {\em Solutions of ill-posed problems},
\newblock Scripta series in mathematics. Winston and distributed solely by
  Halsted Press, 1977.

\bibitem{music}
R.~Schmidt,
\newblock ``Multiple emitter location and signal parameter estimation,''
\newblock {\em IEEE Transactions on Antennas and Propagation}, vol. 34, no. 3,
  pp. 276--280, 1986.

\bibitem{DPD_Nadiri}
O.~Nadiri and B.~Rafaely,
\newblock ``Localization of multiple speakers under high reverberation using a
  spherical microphone array and the direct-path dominance test,''
\newblock {\em IEEE/ACM Transactions on Audio, Speech, and Language
  Processing}, vol. 22, no. 10, pp. 1494--1505, 2014.

\bibitem{image_method}
J.~Allen and D.~Berkley,
\newblock ``Image method for efficiently simulating small-room acoustics,''
\newblock {\em The Journal of the Acoustical Society of America}, vol. 65, pp.
  943--950, 04 1979.

\bibitem{timit}
J.~S. Garofolo,
\newblock ``Timit acoustic phonetic continuous speech corpus,''
\newblock {\em Linguistic Data Consortium, 1993}, 1993.

\bibitem{LOCATA}
H.~W. Löllmann, C.~Evers, A.~Schmidt, H.~Mellmann, H.~Barfuss, P.~A. Naylor,
  and W.~Kellermann,
\newblock ``The locata challenge data corpus for acoustic source localization
  and tracking,''
\newblock in {\em 2018 IEEE 10th Sensor Array and Multichannel Signal
  Processing Workshop (SAM)}, 2018, pp. 410--414.

\end{thebibliography}

\end{document}